\documentclass[]{aastex63}

\received{}
\revised{}
\accepted{}
\submitjournal{ApJ}

\shorttitle{CH$_4$ formation with H$_2$}
\shortauthors{Lamberts et al.}
\watermark{DRAFT}
\graphicspath{{./}{figures/}}

\usepackage{graphicx}
\usepackage[version=4]{mhchem}
\usepackage{units}
\usepackage{txfonts}
\usepackage{amsmath,amssymb}
\usepackage{xcolor} 

\begin{document}

\title{Methane formation in cold regions from carbon atoms and molecular hydrogen}

\correspondingauthor{Thanja Lamberts}
\email{a.l.m.lamberts@lic.leidenuniv.nl}

\author[0000-0001-6705-2022]{Thanja Lamberts}
\affiliation{Leiden Institute of Chemistry, Gorlaeus Laboratories, Leiden University, PO Box 9502, 2300 RA Leiden, The Netherlands}
\affiliation{Laboratory for Astrophysics, Leiden Observatory, Leiden University, PO Box 9513, 2300 RA Leiden, The Netherlands}

\author[0000-0003-2434-2219]{Gleb Fedoseev}
\affiliation{Laboratory for Astrophysics, Leiden Observatory, Leiden University, PO Box 9513, 2300 RA Leiden, The Netherlands}
\affiliation{Research Laboratory for Astrochemistry, Ural Federal University, Kuibysheva St. 48, 620026 Ekaterinburg, Russia}

\author{Marc van Hemert}
\affiliation{Leiden Institute of Chemistry, Gorlaeus Laboratories, Leiden University, PO Box 9502, 2300 RA Leiden, The Netherlands}

\author[0000-0002-3276-4780]{Danna Qasim}
\affiliation{Laboratory for Astrophysics, Leiden Observatory, Leiden University, PO Box 9513, 2300 RA Leiden, The Netherlands}
\affiliation{Current address: Astrochemistry Laboratory, NASA Goddard Space Flight Center, Greenbelt, MD 20771, USA}

\author[0000-0001-6877-5046]{Ko-Ju Chuang}
\affiliation{Laboratory for Astrophysics, Leiden Observatory, Leiden University, PO Box 9513, 2300 RA Leiden, The Netherlands}

\author[0000-0002-3401-5660]{Julia C. Santos}
\affiliation{Laboratory for Astrophysics, Leiden Observatory, Leiden University, PO Box 9513, 2300 RA Leiden, The Netherlands}

\author[0000-0002-8322-3538]{Harold Linnartz}
\affiliation{Laboratory for Astrophysics, Leiden Observatory, Leiden University, PO Box 9513, 2300 RA Leiden, The Netherlands}

\begin{abstract}
\noindent Methane is typically thought to be formed in the solid state on the surface of cold interstellar icy grain mantles \emph{via} the successive atomic hydrogenation of a carbon atom. In the current work we investigate the potential role of molecular hydrogen in the \ce{CH4} reaction network. We make use of an ultra-high vacuum cryogenic setup combining an atomic carbon atom beam and both atomic and/or molecular beams of hydrogen and deuterium {on a \ce{H2O} ice}. These experiments lead to the formation of methane isotopologues detected \emph{in situ} through reflection absorption infrared spectroscopy. Most notably, \ce{CH4} is formed in an experiment combining C atoms with \ce{H2} on amorphous solid water, albeit slower than in experiments with H atoms present. Furthermore, \ce{CH2D2} is detected in an experiment of C atoms with \ce{H2} and \ce{D2} on \ce{H2O} ice. \ce{CD4}, however, is only formed when D atoms are present in the experiment. These findings have been rationalized by means of computational chemical insights. This leads to the following conclusions: a) the reaction \ce{C + H2 -> CH2} can take place, although not barrierless {in the presence of water}, b) the reaction \ce{CH + H2 -> CH3} is barrierless, but has not yet been included in astrochemical models, c) the reactions \ce{CH2 + H2 -> CH3 + H} and \ce{CH3 + H2 -> CH4 + H} can take place only \emph{via} a tunneling mechanism {and d) molecular hydrogen possibly plays a more important role in the solid-state formation of methane than assumed so far.}
\end{abstract}



\section{Introduction} \label{sec:intro}

Methane, the smallest hydrocarbon, is one of the few molecules that have been detected in the solid phase in various regions in the interstellar medium (ISM) \citep{Boogert:2015}. In fact, the first detection was a simultaneous gas phase and tentative solid phase identification, based on the $\nu_4$ feature at 7.6 $\mu$m \citep{Lacy:1991} and meanwhile several in-depth observational studies have been reported \citep{Boogert:1996,Oberg:2008}. Early reports based on comparison to laboratory data indicated that methane likely resides in ice comprising polar component(s) \citep{Boogert:1996} and it was later postulated that \ce{H2O} is the primary candidate for this based on correlations between \ce{CH4} and \ce{H2O} column densities \citep{Oberg:2008}. This points to the fact that solid \ce{CH4} is formed during the translucent phase of the evolutionary track of molecular clouds. 

The solid-state formation of methane is typically assumed to follow four sequential atomic hydrogenation steps of the carbon atom in the $^{3}$P ground state ever since this was postulated in the late 1940s \citep{Hulst:1946, Hulst:1949, Hendecourt:1985, Brown:1988, Brown:1991}. Recently, this route has been confirmed experimentally \citep{Qasim:2020a} through the simultaneous use of well-characterized C- and H-atom beams, following up on early work by \citep{Hiraoka:1998}:
\begin{align}
    \ce{^3C + H} &\ce{-> CH} \label{C+H}\\
    \ce{CH + H} &\ce{-> CH2} \label{CH+H} \\
    \ce{^3CH2 + H} &\ce{-> CH3} \label{CH2+H}\\
    \ce{CH3 + H} &\ce{-> CH4} \label{CH3+H}\;.
\end{align}

The hydrogen atom number density in molecular clouds is estimated to be around a few atoms/cm$^{3}$, which is two to four orders of magnitude lower than the molecular hydrogen abundance depending on whether a translucent or dense cloud is concerned \citep{Dishoeck:1988,Goldsmith:2005}. Therefore, also solid-state reactions with molecular hydrogen can be of great importance even when the corresponding rate constants are lower, as pointed out already in 1993 by \citeauthor{Hasegawa:1993}. For instance for the sequential hydrogenation of the O atom to eventually form water \citep{Hiraoka:1998, Ioppolo:2008, Miyauchi:2008} it has been shown that the reaction \ce{H2 + OH -> H2O + H} can become more relevant than \ce{H + OH -> H2O} even though a considerable barrier is invoked \citep{Cuppen:2007,Furuya:2015}. Another example is the hydrogenation of carbon monoxide by UV-irradiation of mixed CO:\ce{H2} ices \citet{Chuang:2018}. Despite such active involvement of \ce{H2} in solid-state reactions, the molecular hydrogen abundances on ice surfaces in astrochemical microscopic models are often artificially reduced to save (a lot of) computational cost, \emph{e.g.}, by decreasing the sticking coefficient \citep{Lamberts:2013,Vasyunin:2013,Garrod:2013}.
We also want to point out that despite the fact that neutral carbon atoms are often thought to only be (abundantly) present in translucent regions \citep{Dishoeck:1988, Snow:2006}, there exists substantial literature that leads to believe that C[I] is more extended, possibly even into denser regions \citep{Langer:1976,Keene:1985,Papadopoulos:2004,Burton:2015,Bisbas:2019}.

For these reasons, we consider a number of reactions with molecular hydrogen in the context of methane formation. Firstly, the direct addition or insertion reactions:
\begin{align}
    \ce{^3C + H2} &\ce{-> ^3CH2} \label{C+H2}\\
    \ce{CH + H2} &\ce{-> CH3} \label{CH+H2}\;.
\end{align}
Reaction~\ref{C+H2}, the first step in the reaction network, has been covered in a series of papers indicating that the reaction may readily take place in helium droplets \citep{Krasnokutski:2016, Henning:2019} and including this reaction in astrochemical models without a barrier was suggested recently \citep{Simoncic:2020}. The reaction
\begin{align}
    \ce{^1CH2 + H2} &\ce{-> CH4} \label{notpossible}
\end{align} 
intuitively the most likely step to form methane, can only take place if methylene is in the excited singlet state, \ce{^1CH2} \citep{Murrell:1973, Bauschlicher:1977}. In the system currently under study, only ground-state \ce{^3CH2} is expected to be present and therefore reaction~\ref{notpossible} will not be further considered. 

In terms of the chemical reaction network, there is furthermore the possibility of hydrogen abstraction, either from \ce{H2} or from a \ce{CH_n} fragment:
\begin{align}
\ce{CH + H} & \ce{-> ^3C + H2} \label{C+H2a} \\
\ce{CH2 + H} & \ce{-> CH + H2} \label{CH+H2a} \\
\ce{^3CH2 + H2} & \ce{-> CH3 + H} \label{CH2+H2a} \\
\ce{CH3 + H2} & \ce{-> CH4 + H} \label{CH3+H2a}
\end{align}
{Reactions~\ref{C+H2a}-\ref{CH3+H2a} are listed in the exothermic direction, and the reverse endothermic reactions are not expected to be important given the low temperatures involved ($\sim10-20$~K).}
We assume that the exothermicities of the reactions are not significantly altered by the solid-state surroundings of the reaction site. This essentially translates into assuming that the binding energies of the reactants and products are similar.
{We also assume that the H atoms formed in reactions~\ref{CH2+H2a} and~\ref{CH3+H2a} immediately desorb, based on the argument of conservation of energy and momentum \citep{Koning:2013}.}

In the current manuscript, we revisit reaction~\ref{C+H2} in water ices and extend the discussion on \ce{H2} reactivity investigating the influence of reactions~\ref{CH+H2}--\ref{CH3+H2a}. {We show that reactions~\ref{C+H2}, \ref{CH+H2}, \ref{CH2+H2a}, and~\ref{CH3+H2a} together can lead to the formation of methane, without the involvement of H-atoms. To achieve this,} we make use of in-situ infrared spectroscopy to probe which methane isotopologues are formed via reaction of \ce{^{3}C} with selected combinations of atomic and molecular hydrogen (H, \ce{H2}) and deuterium (D, \ce{D2}) on representative interstellar water-rich ice analogs. 

This manuscript is organized in the following way. The experimental details and results are discussed in sections~\ref{sec:methodology} and~\ref{sec:results}. Furthermore, we position the chemical network listed above within the context of the extensive theoretical chemical literature, {complemented by additional computations presented that explicitly take into account the role of the water ice surface}. This helps to disentangle which reactions are likely to take place throughout the various experiments, as outlined in Section~\ref{sec:theomethod} and discussed in section~\ref{sec:tcresults}.
{Solid methane is hard to observe from ground-based observatories because of telluric pollution. Observations from space offer an alternative. Solid methane has been observed already with the Spitzer space telescope \citep{Oberg:2008}. Because of its higher sensitivity and spatial resolution, the James Webb Space Telescope (JWST) is expected to substantially extend on these observations. The present experimental work and theoretical approach fits worldwide efforts to prepare for upcoming JWST observations. In section \ref{sec:astroconc} the astrochemical implications and conclusions of this work are presented.  }

\begin{table*}
    \centering 
\caption{Summary of all performed experiments, organized along four selected sets. MDL, SUKO and MWAS refer to the used atomic and molecular deposition lines (see text). Furthermore the substrate temperature, the atomic and molecular fluxes and the total time of the experiment, from which the fluence can be derived, are listed. All fluxes give the effective value. Please note that H (\ce{H2}) and D (\ce{D2}) have different thermal velocities and sticking coefficients, thus a direct comparison should be done with care. Values in bold font underline the different settings within one series of measurements.}\label{tab:listexp}
    \begin{tabular}{lrrrrrrrlr}
    \hline
    \# & T   & \ce{H2O}   & C     & H     & \ce{H2}   & D     & \ce{D2} &     C:H:\ce{H2}:D:\ce{D2} & Time \\
   &    K & cm$^{-2}$ s$^{-1}$      & cm$^{-2}$ s$^{-1}$      & cm$^{-2}$ s$^{-1}$      & cm$^{-2}$ s$^{-1}$      & cm$^{-2}$ s$^{-1}$      & cm$^{-2}$ s$^{-1}$     & & (min)\\
    \hline \hline
   &     & MDL      & SUKO      & \multicolumn{2}{c}{MWAS}      & & {MDL}   & & \\

    \hline
    1 A  & 10 & $8\times10^{12}$ & $5\times10^{11}$ & $2\times10^{12}$ & $1\times10^{14}$  &  &  & 1:4:200:--:-- & 30 \\
    1 B  & 10 & $8\times10^{12}$ & $5\times10^{11}$ & $2\times10^{12}$ & $1\times10^{14}$  &  &  $\bf 1\times10^{13}$  & 1:4:200:--:20 & 30 \\    
    1 C  & 10 & $8\times10^{12}$ & $5\times10^{11}$ & $2\times10^{12}$ & $1\times10^{14}$  &  &  $\bf 4\times10^{13}$  & 1:4:200:--:80 & 30 \\    
    \hline
  &      & MDL      & SUKO      &  & {MDL} & \multicolumn{2}{c}{MWAS}  & & \\
    \hline
    2 A  & 10 & $8\times10^{12}$ & $5\times10^{11}$ &  &  & $1.5\times10^{12}$ & $6\times10^{13}$ & 1:--:--:3:120 & 30 \\
    2 B  & 10 & $8\times10^{12}$ & $5\times10^{11}$ &  & $\bf 5\times10^{13}$  & $1.5\times10^{12}$ & $6\times10^{13}$ & 1:--:100:3:120 & 30 \\
    2 C  & 10 & $8\times10^{12}$ & $5\times10^{11}$ &  & $\bf 2\times10^{14}$  & $1.5\times10^{12}$ & $6\times10^{13}$ & 1:--:400:3:120 & 60 \\    
    \hline
 & & MDL      & SUKO      & & {MDL} & & MDL   & & \\
    \hline
    3 A  & 10 & $1.2\times10^{13}$ & $5\times10^{11}$ & & & & $\bf 1\times10^{14}$ &  1:--:--:--:200 & 60 \\ 
    3 B  & 10 & $1.2\times10^{13}$ & $5\times10^{11}$ & & $\bf 1\times10^{14}$ & & $\bf 1\times10^{14}$ &  1:--:200:--:200 & 240 \\
    \hline  \hline
    \# & T & \ce{D2O}   & C     &  & \ce{H2}   &      &  &  &  Time \\
   &    K & cm$^{-2}$ s$^{-1}$      & cm$^{-2}$ s$^{-1}$      & cm$^{-2}$ s$^{-1}$      & cm$^{-2}$ s$^{-1}$      & cm$^{-2}$ s$^{-1}$      & cm$^{-2}$ s$^{-1}$     & & (min)\\
\hline \hline
 & & MDL      & SUKO      &  & {MDL}  & &  & & \\
    \hline
    4 A  & $\bf 10$ & $1.4\times10^{13}$ & $5\times10^{11}$ & & $ 2.5\times10^{14}$ & & & 1:--:500:--:-- & 60 \\
    4 B  & $\bf 25$ & $1.4\times10^{13}$ & $5\times10^{11}$ & & $ 2.5\times10^{14}$ & & & 1:--:500:--:-- & 60 \\
    \hline
    \end{tabular}
\end{table*}

\section{Methodology} \label{sec:methods}

\subsection{Experimental methodology} \label{sec:methodology}

The experimental setup used is SURFRESIDE$^3$, an ultra-high vacuum system with three atomic beam lines \citep{Ioppolo:2013, Qasim:2020b}. For the purpose of our study, only H/D and C atom beam lines are used. Ices are grown on a gold-coated copper substrate that is attached to the cold finger of a closed-cycle He cryostat in the centre of the main UHV chamber with a base pressure on the order of 10$^{-10}$ mbar. Co-deposition experiments of \ce{H2O + ^3C} and combinations of H, D, \ce{H2}, and/or \ce{D2} are performed that lead to the growth of a mixed ice at 10 K. Mixed \ce{H/H2}- or \ce{D/D2}-beams are obtained by (partial) dissociation of { molecular \ce{H2} (Linde 5.0) or \ce{D2} (Linde 2.8) }in a microwave discharge source (MWAS, Oxford Scientific, \citep{Schmidt:1996, Anton:2000}) in a separate vacuum chamber with a base pressure of $\sim$10$^{-9}$ mbar. Note that charged particles are removed by means of applying an electric field that deflects these species. Excited-state species are de-excited through collisions with the walls of a U-shaped quartz pipe at room temperature placed along the beam path prior to the molecules entering the main chamber.
A customized SUKO-A 40 C-atom source \citep{Qasim:2020b} based on a commercial design (Dr. Eberl, MBE, \citep{Krasnokutski:2014, Albar:2017}) produces a beam of carbon atoms in the $^3$P ground state with a \ce{C_n}/C (n$>$1) ratio of less than 0.01. This source is located in another separate vacuum chamber with a base pressure (3-5)$\times$10$^{-9}$ mbar. A series of apertures is used to collimate the C-atom beam on the substrate avoiding deposition of carbon on the walls of the main UHV chamber. Note that CO and \ce{CO2} are known (minor) contaminants in the C-atom beam. A third vacuum chamber with a base  pressure of $\sim$10$^{-9}$ mbar is used to generate a molecular { \ce{H2O} (Milli-Q) or \ce{D2O} ( Sigma-Aldrich 99.9 atom\% D) beam for simultaneous deposition (MDL).} We use an overabundance of \ce{H2O}, to mimic a polar ice environment, \emph{e.g.}, in agreement with the previously mentioned correlation between observed \ce{CH4} and \ce{H2O} column densities. 
The atomic and molecular fluxes are listed in Table~\ref{tab:listexp}, in which all experiments are summarized. The uncertainty in the flux determination is roughly a factor two.
The initial reactants and formed products are monitored via Reflection Absorption InfraRed Spectroscopy (RAIRS, \citep{Greenler:1966}) in the solid state \emph{in situ}. These experiments are organized along four different series in which relevant parameters are systematically varied. 

\subsection{Computational methodology}\label{sec:theomethod}

\begin{figure*}
\centering
\includegraphics[width=0.8\textwidth]{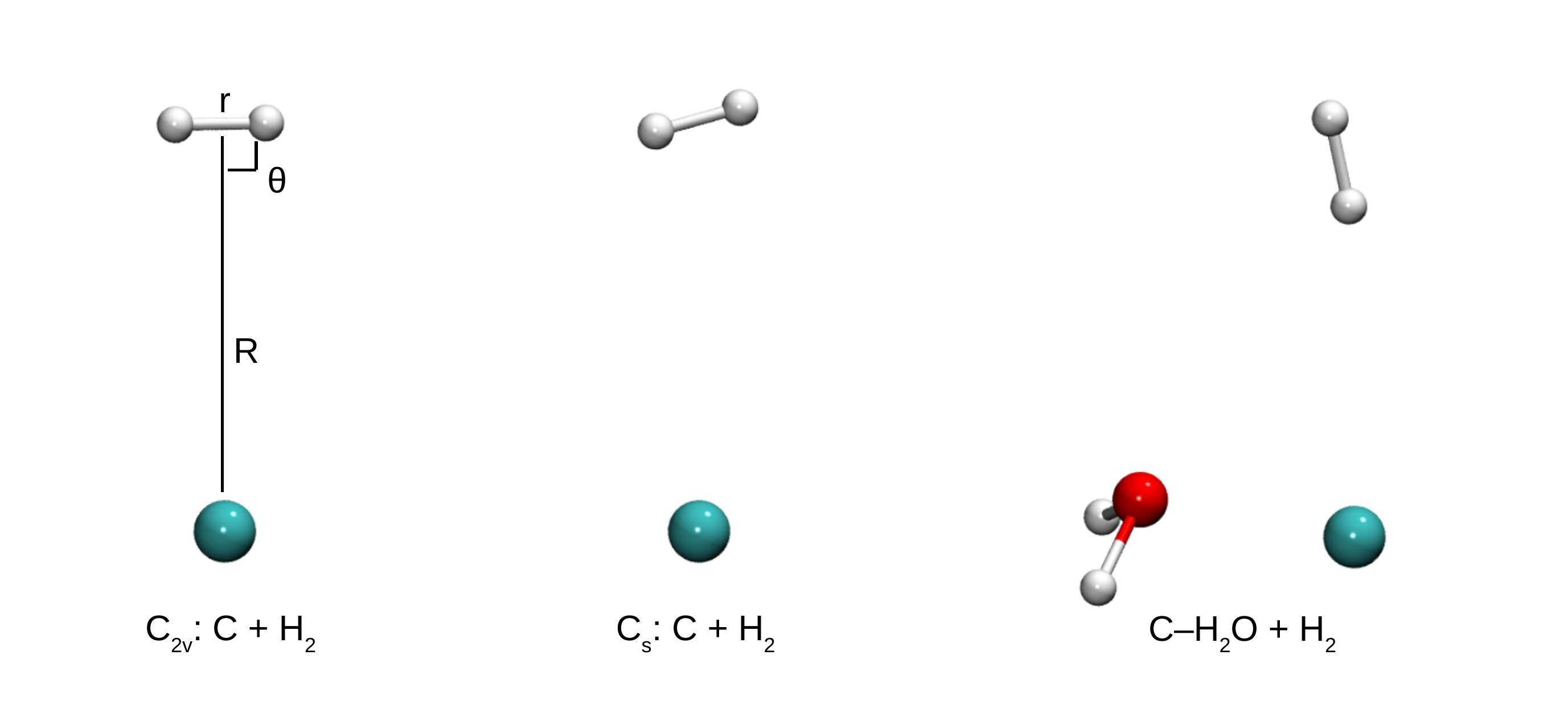}
\caption{Three cases studied for the reaction \ce{C + H2 -> CH2}.}\label{fig:C+H2}
\end{figure*}

The primary focus of our own calculations is on the reaction \ce{C + H2 -> CH2}, since it is the first and most determining step in the reaction network mentioned above leading to \ce{CH4}. Moreover, the carbon atom is known to be highly reactive \citep{Kim:2003}. However, despite the work by \citet{Simoncic:2020}, to date it has not been looked at it in detail for surface chemistry purposes. Note that for the other possible steps (reactions~\ref{CH+H2}--\ref{CH3+H2a}), we draw from previous studies available from the literature.

We take a threefold approach, see also Fig.~\ref{fig:C+H2}, to understand the reactivity of \ce{C + H2 -> CH2}: 
\begin{enumerate}
\setlength{\itemsep}{0pt}
\setlength{\parskip}{0pt}
    \item a gas-phase calculation of a highly symmetric orientation of C with respect to \ce{H2} (\ce{C_{2v}} symmetry) - meant to compare against previous results,
    \item a gas-phase calculation of a low symmetry orientation of C with respect to \ce{H2} (\ce{C_{s}} symmetry) - meant to be a first step towards a realistic symmetry-broken orientation on a surface,
    \item a calculation of the reaction of \ce{H2} with a C atom bound to a single \ce{H2O} molecule (no symmetry) - meant as a first step towards understanding the reaction with a bound carbon atom on a water ice.
\end{enumerate}

When an \ce{H2} molecule approaches a carbon atom in its \ce{^3P} ground state the encounter can occur on three, initially degenerate, potential surfaces (PESs). Important to take into account is the alignment of the two singly occupied 2p orbitals of the carbon atom. When the orientation of the \ce{H2} molecule is perpendicular to the direction of approach, there is a \ce{C_{2v}} symmetry and the three surfaces are \ce{^3B1}, \ce{^3B2} and \ce{^3A2}. For all three surfaces the energy was calculated as a function of the Jacobi coordinates $R$ as the distance between C and the center-of-mass of \ce{H2}, $r$ as the H--H distance, and $\theta=90^\circ$ as the angle between $R$ and $r$, see again Fig.~\ref{fig:C+H2}. All calculations were repeated in \ce{C_s} symmetry where the angle $\theta$ was reduced to 80$^\circ$, while keeping the CAS/CI parameters as closely as possible to the ones used in the \ce{C_{2v}} case. In \ce{C_s} symmetry two \ce{^3A}" states are relevant. Subsequently, to find possible reaction paths, the location of the crossing (seem) between the \ce{^3B1} and \ce{^3A2} for \ce{C_{2v}} and the two \ce{^3A}" surfaces for \ce{C_s} was searched for by scanning these surfaces with small steps in the $R\approx r \approx 1$ $\AA$ range. These calculations have been performed using Molpro \citep{Molpro, molpro2018} with the AVQZ basis set \citep{Woon:1993,Dunning:2001}. 
The CASSCF calculation had 8 electrons in 8 orbitals, while state averaging over singlet and triplet states, each with 4 lowest roots, both for \ce{C_{2v}} and \ce{C_s} symmetries.

In order to test -- for the first time -- the influence of {the strong adsorption of the carbon atom atop water ice} on the reactivity of the carbon atom with \ce{H2} we have performed a nudged elastic band (NEB) calculation for the reaction of \ce{H2} with a carbon atom bound to a water molecule, \emph{i.e.}, \ce{C-H2O} complex. This was followed by a transition state optimization applying the dimer method with the B3LYP functional \citep{Becke:1988, Becke:1993, Lee:1988} and using DL-find in Chemshell \citep{DLfind,Chemshell}. For the resulting reactant, transition, and product states, single-point energies were calculated with both {MRCI/AVTZ} \citep{MRCI, Dunning:1989, Kendall:1992} and CCSD(T)-F12a/VTZ-F12 \citep{CCSDTF12_2007, CCSDTF12_2009, Peterson:2008} in Molpro version 2020 \citep{molpro2020}. All geometries were re-optimized at the CCSD(T)-F12a/VTZ-F12 level of theory in Molpro. MRCI calculations are, in principle, warranted for systems where multi-reference effects can be expected, such as here where a triplet ground state carbon atom is involved. We do note, however, that the $T1$ and $D1$ values in both CCSD(T)-F12 calculations are (well) below the common threshold values of {$T1 <0.04$ and $D1 <0.05$} \citep{Janssen:1998, Lambert:2006}. Indeed, the MRCI single-point energy calculations indicate that the main contribution of a Slater determinant to the wave function has {reference coefficients of about 0.93-0.94} for all \ce{C - H2O - H2} geometries.

\begin{figure*}
\includegraphics[width=1\textwidth]{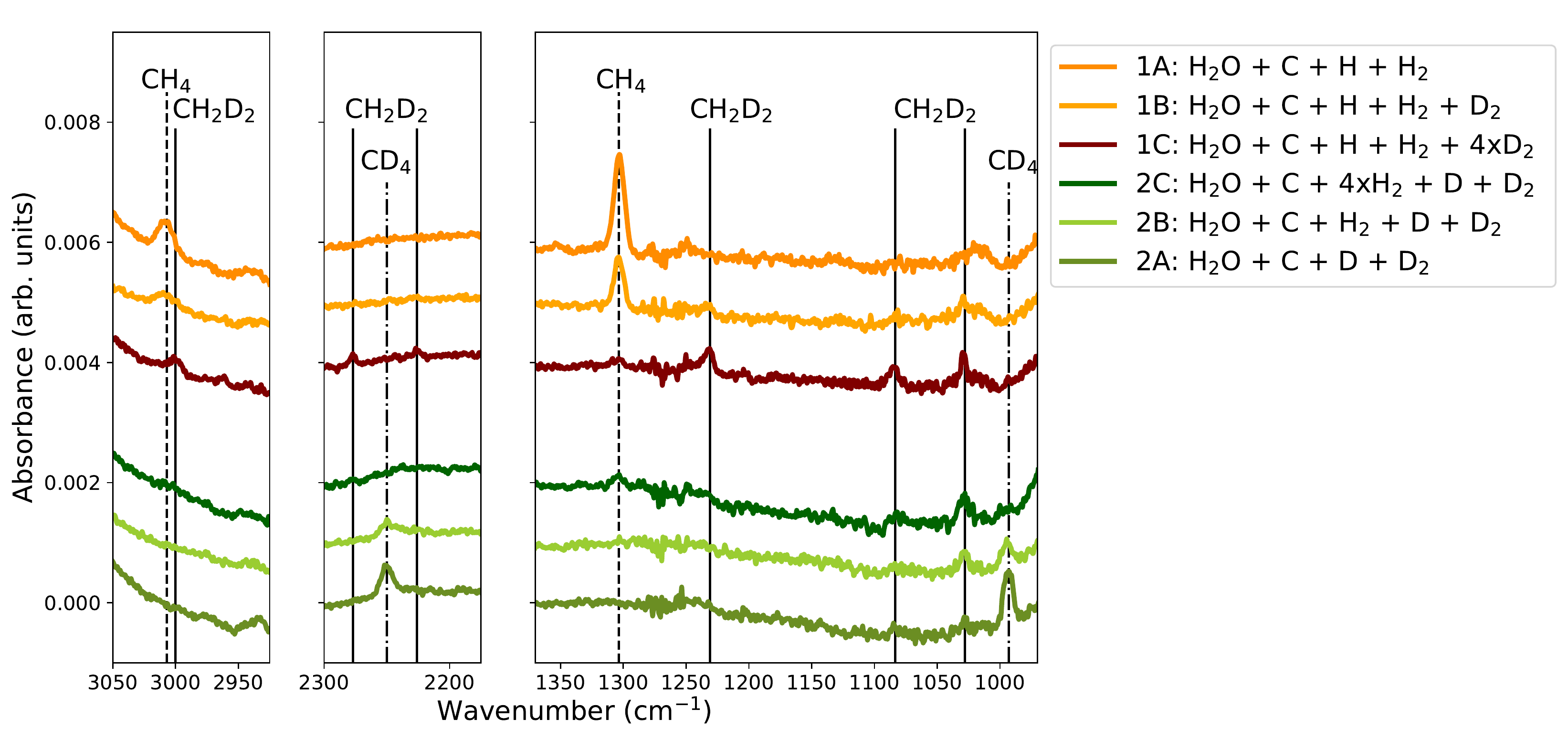}
\caption{RAIR spectra of experimental series~1 and~2, the exact experimental conditions can be found in Table~\ref{tab:listexp}. The dashed vertical lines indicate the peak positions of \ce{CH4}, the solid vertical lines indicate \ce{CH2D2}, the dash-dotted lines indicate \ce{D2CO}, see Table~\ref{tab:peaks} in the appendix for specific peak positions.}\label{fig:fullspec}
\end{figure*}

\section{Experimental results} \label{sec:results}

Figure~\ref{fig:fullspec} shows the RAIR spectra of experimental series 1 and 2, over the wavenumber ranges that include all relevant spectroscopic bands of \ce{CH4}, \ce{CD4}, and \ce{CH2D2}, indicated by the vertical lines in the figure. {Other detected species are listed in Table~\ref{tab:peaks} in Appendix~\ref{sec:peaks}.}

Experiments 1A and 2A (NB: top and bottom graph in Fig.~\ref{fig:fullspec}) serve as a control experiment and can be directly compared to previous work \citep{Qasim:2020a}: indeed the formation of \ce{CH4} and \ce{CD4} is clearly confirmed by the presence of both the $\nu_3$ and $\nu_4$ vibrational modes. Molecular deuterium (hydrogen) is introduced in experiment 1B (2B) and increased in experiment 1C (2C) and a concomitant decrease of \ce{CH4} (\ce{CD4}) can be easily observed. At the same time, several \ce{CH2D2} absorption features appear in these four experiments, the most intense peak at 1028 cm$^{-1}$. Experiment 1C shows the most apparent, multi-line detection of doubly deuterated methane as a result of the advantageous ratio between all reactants, \emph{i.e.}, C:H:\ce{H2}:\ce{D2} = 1:4:200:80, see also Table~\ref{tab:listexp}. 

Because doubly deuterated methane is observed in experiments for which either hydrogen or deuterium is present only in the molecular form, at least one reaction with a molecular species must take place throughout the course of methane formation. This is further supported by a tentative \ce{CH4} detection at 1303 cm$^{-1}$ in experiment 2C, \emph{i.e.}, in an experiment with hydrogen present only in the molecular form.

\begin{figure*}
    \centering
    \includegraphics[width=0.75\textwidth]{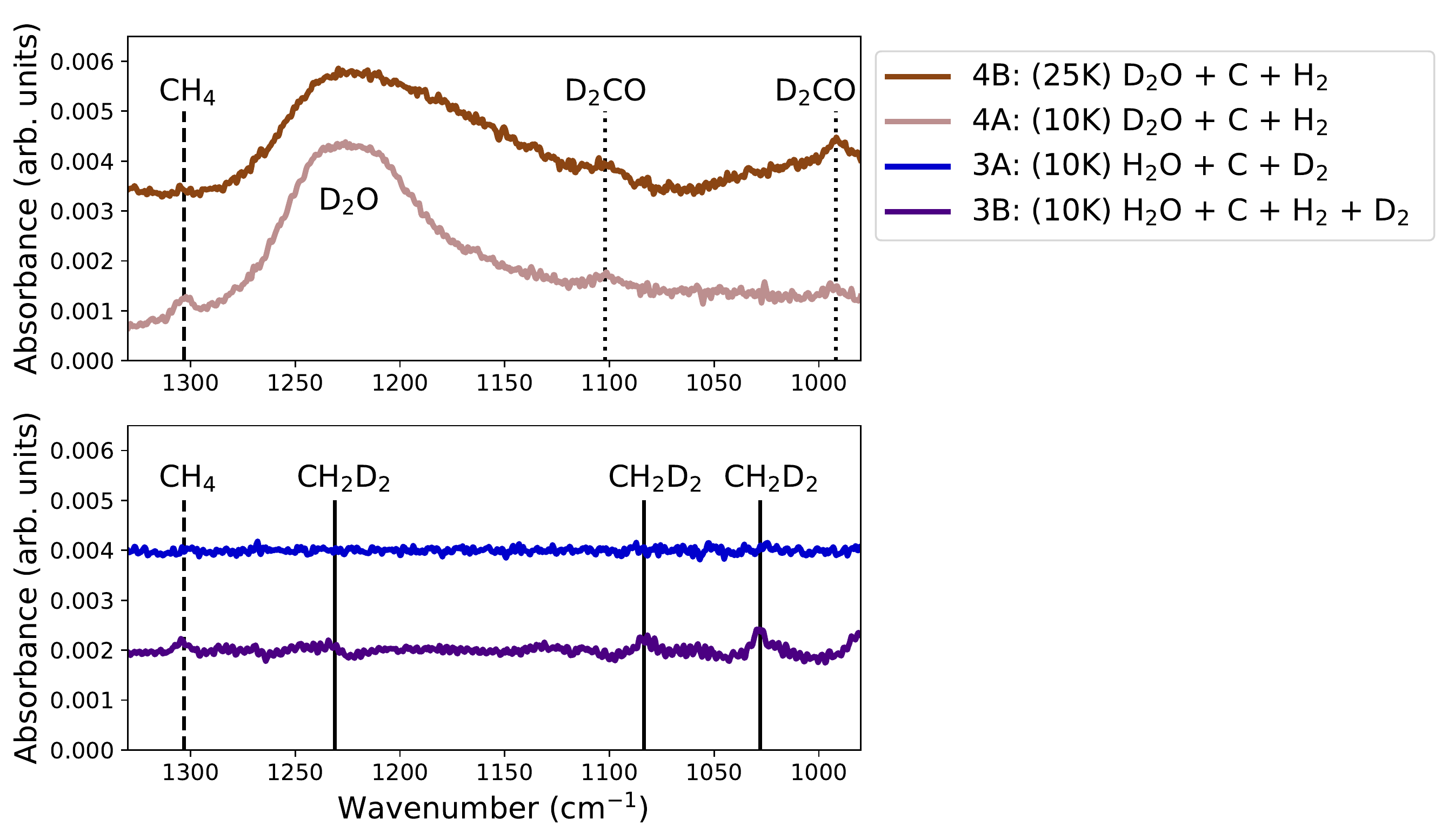}
    \caption{RAIR spectra of experiments 1C, 2C, 3A, 3B, 4A, and 4B, the exact experimental conditions can be found in Table~\ref{tab:listexp}. The dashed vertical lines indicate the peak position of \ce{CH4}, the solid vertical lines indicate \ce{CH2D2}, the dotted lines indicate \ce{D2CO}, and the broad peak is caused by \ce{D2O} indicated directly in the plot, see Table~\ref{tab:peaks} in the appendix for specific peak positions.} \label{fig:zoomspec}
\end{figure*}

This tentative detection was the reason to perform experimental series 3 and 4 to further investigate whether methane indeed can be formed without the presence of atomic species. In order to increase the signal-to-noise ratio, these experiments have been run over longer times, as indicated in Table~\ref{tab:listexp}. {These experiments clearly result in a lower overall formation rate of methane in comparison to interactions with atomic species.
The resulting thick ices ($\sim 200$~ML)} and concomitant interference patterns over the full range of the IR spectra have led to presenting baseline-corrected spectra of the region of interest only ($1330 - 980$ cm$^{-1}$) depicted in Fig.~\ref{fig:zoomspec}. {The baseline-correction includes the subtraction of known components of the purging gas used along the path of the infrared beam, \emph{i.e.}, outside the vacuum chamber.} Experiment 3B, consisting of C + \ce{H2} + \ce{D2} on \ce{H2O} ice, shows a clear detection of the two main \ce{CH2D2} peaks at 1083 and 1028 cm$^{-1}$. In other words, \ce{CH2D2} has been formed from carbon atoms and molecular species only. {However, the yield of \ce{CH2D2} is significantly less than in experiments 1C and 2C.} Experiments 3A and 4A represent a co-deposition of C + \ce{D2} + \ce{H2O}, and C + \ce{H2} + \ce{D2O}, respectively. In experiment 3A no \ce{CD4} is detected, while experiment 4A shows a clear solid-state \ce{CH4} detection. In other words in the reaction network for the formation of methane from carbon atoms and molecular hydrogen there is an isotope effect present, which hints for the importance of tunneling in one, or more, of the involved reactions. 

Experiment 4B, another control experiment, performed at 25 K, \emph{i.e.}, above the desorption temperature of atomic or molecular hydrogen, shows no \ce{CH4} formation in the solid state confirming that the processes we are studying take place at the surface at low temperatures. 

{Finally, no \ce{CD3H} or \ce{CH3D} was detected in any of the experiments which can be attributed to a combination of isotope effects, lower band strengths, and competition between tunneling and barrierless reactions in the network. A full rationale for the lack of these signals is given in Appendix~\ref{sec:triple}. }

We can summarize the experimental findings as follows:
\begin{enumerate}
  \setlength{\itemsep}{0pt}
  \setlength{\parskip}{0pt}
    \item \ce{CH4} is detected in experiments on a water ice using both H atoms and \ce{H2} molecules as well as in experiments using only \ce{H2}, albeit with a lower efficiency,
	\item \ce{CD4} is detected only in experiments where D atoms are present,
	\item \ce{CH2D2} can be detected in experiments that use simultaneously both atomic and molecular hydrogen or deuterium sources,
	\item \ce{CH2D2} can also be detected in experiments that use exclusively molecular \ce{H2} and \ce{D2},
	\item Neither \ce{CH3D} nor \ce{CHD3} are detected,
	\item All studied reactions take place on the cold ice surface.
\end{enumerate}

\section{Theoretical results}\label{sec:tcresults}

\begin{figure*}
    \centering
    \includegraphics[width=12.5cm]{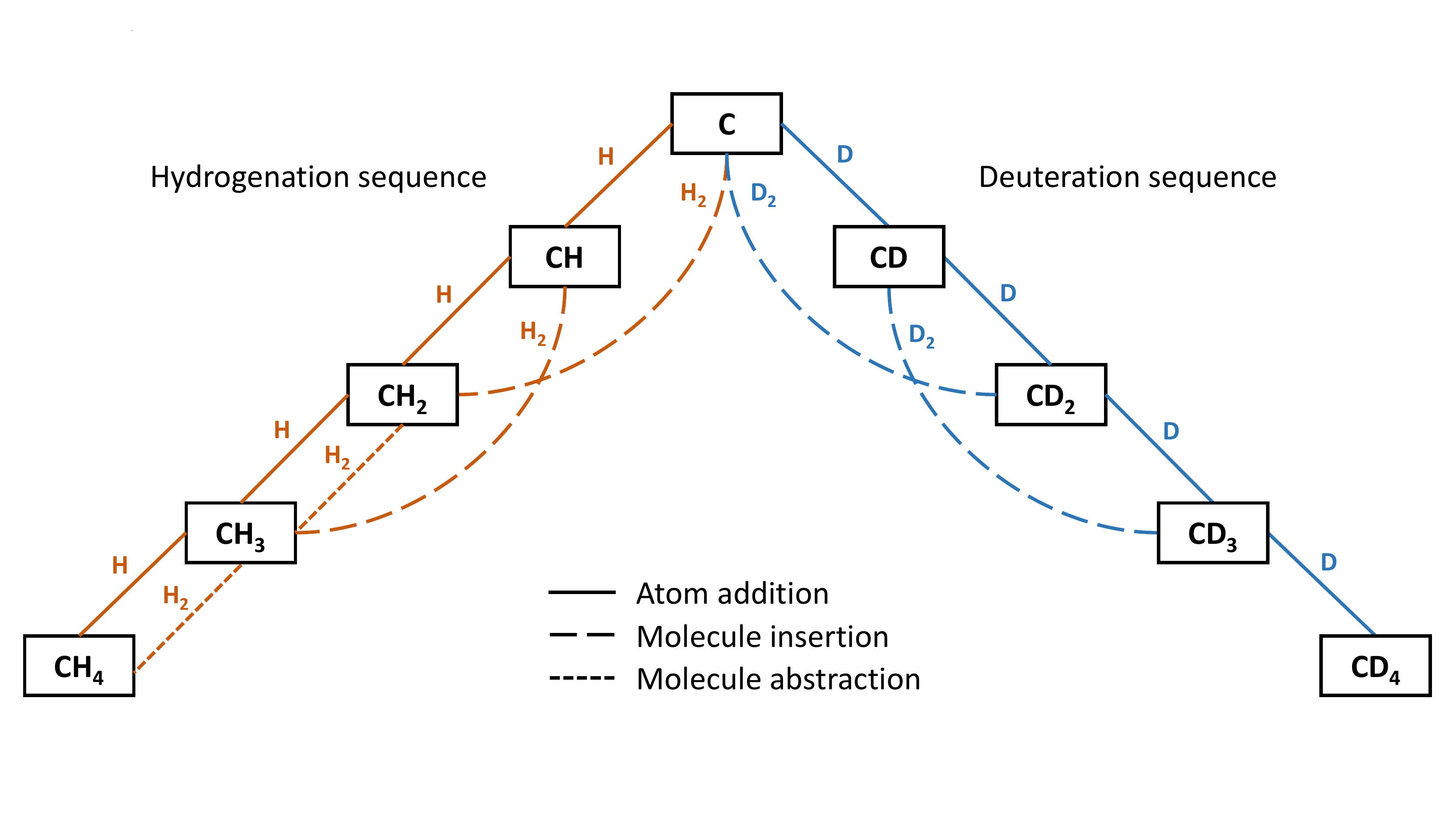}
    \caption{Three types of reactions are considered to lead to the formation of \ce{CH4} or \ce{CD4}: H or D addition \citep{Qasim:2020a}, \ce{H2} or \ce{D2} insertion (\citet{Krasnokutski:2016, Simoncic:2020} and this work), and \ce{H2} abstraction (this work). We assume that the latter is not efficient with \ce{D2}, based on the involved high barrier and lack of tunneling efficiency.}
    \label{fig:network}
\end{figure*}

This section focuses on explaining the experimental results by looking in detail at theoretical chemical studies of reactions~\ref{C+H2}--\ref{CH3+H2a}. Reactions~\ref{C+H}--\ref{CH3+H} have been extensively discussed in \citet{Qasim:2020a} and will therefore not be further dealt with here. {A summary of all relevant reaction steps can be found in Fig.~\ref{fig:network}}

Section~\ref{sec:CH2PES} contains predominantly results from our own calculations, while sections~\ref{sec:CH3PES} and~\ref{sec:AbstrRxn} are based on previous results available from the literature.

\subsection{Reactions on the {CH$_2$} PES}\label{sec:CH2PES}
All three calculated \ce{C_{2v}} surfaces for the \ce{C + H2} reaction, \ce{^3B1}, \ce{^3B2} and \ce{^3A2} are degenerate for large $R$ (C--\ce{H2} distance) with an energy of $\sim$330 kJ/mol above the \ce{^3B1} \ce{CH2} ground state, see also Fig.~\ref{fig:pes} in Appendix~\ref{sec:pes}. 
We find a crossing between the \ce{^3A2} and \ce{^3B1} surfaces at $R=1.05~\AA$ and $r=1.17~\AA$ at an energy equal to the asymptotic value of \ce{C +H2}. In agreement with \citet{Gamallo:2012}, under the assumption that a transition from the \ce{^3A2} to the \ce{^3B1} occurs via a conical intersection, there is no barrier.
When the symmetry is lowered to \ce{C_s}, which should more closely resemble the situation on an ice surface, by giving up the perpendicular orientation and instead choosing $\theta=80^\circ$, both the  \ce{^3B1} and \ce{^3A2} surfaces become of the same \ce{^3A}" symmetry and a reaction can take place without a serious energetic barrier. A small hump can be found along the path at $R=1.08~\AA$ and $r=1~\AA$, but this remains below the \ce{C + H2} asymptotic value, see Fig.~\ref{fig:pes}.

\begin{table*}
\centering
\caption{Activation and reaction energies in kJ/mol, $E_\text{act}$ and $E_\text{react}$ respectively, for the reaction of the C--\ce{H2O} complex with \ce{H2} leading to the \ce{CH2}--\ce{H2O} complex. The interaction energy, $E_\text{int}$ of the \ce{CH2}--\ce{H2O} complex is also given.}\label{tab:ch2h2o}
\begin{tabular}{lcccc}
\hline
       & B3LYP & MRCI / B3LYP & CCSD(T)-F12 / B3LYP & CCSD(T)-F12 \\
\hline
    $E_\text{int, \ce{C-H2O}}$ & -52.9 & & & -36.1 \\
    $E_\text{act, \ce{C-H2O + H2}}$ & 30.4 & 30.0 & 27.3 & 30.4 \\
    $E_\text{react, \ce{C-H2O + H2}}$ & -294.3 & -281.5 & -303.1 & -300.5 \\
    \hline
    $E_\text{int, \ce{CH2-H2O}}$ & -8.0 & & & -7.1 \\
\hline
\end{tabular}
\end{table*}

While the notion of the reaction \ce{C + H2 -> CH2} being barrierless seems in line with recent literature \citep{Gamallo:2012, Simoncic:2020}, two important things have up to now not been explicitly taken into account. 
First of all, at the (avoided) crossing of the \ce{C_{2v}} potential energy surfaces, there is a change in electron configuration. The dominant contribution to the wavefunction on the \ce{^3A2} surface consists out of an occupation for the 6 valence electrons of (2a$_1$)$^2$ (3a$_1$)$^2$ (1b$_1$)$^1$ (1b$_2$)$^1$ and this needs to change to (2a$_1$)$^2$ (3a$_1$)$^1$ (1b$_1$)$^1$ (1b$_2$)$^2$ before reaching the \ce{^3B_1} \ce{CH2} ground state. This means an electron has to move from a pz orbital on the carbon atom into a py orbital, which can impede the reaction from taking place, even if energetically there is no barrier. This argument still holds in the case of the low-symmetry \ce{C_s} surface where the electron configuration also changes upon decreasing $R$ and increasing $r$, \emph{i.e.}, reacting towards \ce{CH2}.

Secondly, the known strong interaction of \ce{^3P} carbon atom with a \ce{H2O} molecule or cluster \citep{Wakelam:2017,Shimonishi:2018,Duflot:2021,Molpeceres:2021} might drastically change the energetics of the reaction. Estimates on the exact interaction energy depend on the number of water molecules considered and vary between $\sim$35 and $\sim$115 kJ/mol, much larger than a typical water hydrogen bond of some $\sim$22 kJ/mol. Table~\ref{tab:ch2h2o} gives an overview of the interaction energy of the \ce{C-H2O} complex and the activation and reaction energy of the reaction \ce{H2 + C - H2O -> CH2 - H2O} for different levels of theory. 
A sizable barrier of 27-30 kJ/mol is found, which certainly cannot be trivially overcome at 10~K. {This demonstrates that while the \ce{C + H2} reaction proceeds barrierlessly in helium droplets \citep{Krasnokutski:2016, Henning:2019} this is unlikely to be the case on water surfaces. } 
Because this is a proof-of-principle calculation, and more extensive work should be done on C--(\ce{H2O})$_n$ clusters with $n>1$, we cannot elaborate on the expected isotope effect of this reaction, but we assume that this reaction on a water ice proceeds with an (effective) barrier and will thus be slower with \ce{D2} compared to \ce{H2}. 

Reaction~\ref{C+H2a}, \ce{CH + H -> ^3C + H2},  has to proceed through a \ce{CH2} intermediate, as can be seen in Fig.~2 of \citet{Gamallo:2012}. While this is a realistic scenario in gas-phase experiments in the low-pressure regime where intermediates can convert the reaction energy as internal energy to overcome subsequent barriers, such intermediates are expected to be quenched on a surface by rapid ($< 1$~ps) energy dissipation to a water ice \citet{Fredon:2021}. Therefore, reaction~\ref{C+H2a} {turns effectively into} reaction~\ref{C+H2} on a surface.

\subsection{Reactions on the {CH$_3$} PES}\label{sec:CH3PES}
The exchange reaction \ce{CH + H2 <=> CH2 + H} has been previously studied in the gas phase \citep{McIlroy:1993, Medvedev:2006,Gonzalez:2011}. In particular, \citet{Medvedev:2006} showed that the reaction proceeds through an activated \ce{CH3^*} intermediate, which is formed \emph{via} a barrierless pathway in both the forward and backward direction of the reaction, \emph{i.e.}, \ce{CH + H2 <=> CH3^* <=> CH2 + H}. In the low-pressure regime, \citet{Gonzalez:2011} showed that the three hydrogen atoms become equivalent, as a result of the long-lived \ce{CH3^*} complex. In the high-pressure regime, on the other hand, \citet{McIlroy:1993} indicated that collisional stabilization to form ground-state \ce{CH3} dominates. A reaction taking place on an ice surface can be seen as an extreme case of the high-pressure limit, in which the ice acts as a third body to take up the excess energy of the reaction (see above). Given the barrierless nature of the PES, we expect no isotope effect here. In other words, it is highly likely that reaction~\ref{CH+H2} will lead barrierlessly to the formation of \ce{CH3} on the ice, while reaction~\ref{CH+H2a} is unlikely to take place at all and will revert simply to reaction~\ref{CH2+H}, also forming \ce{CH3}. This serves as an explanation why \ce{CH2D2} is detected in experimental sets 1 and 2, since the barrierless nature leads to a lack of isotope effect. 

\subsection{Hydrogen abstraction reactions}\label{sec:AbstrRxn}
The two remaining reactions with molecular hydrogen, reactions~\ref{CH2+H2a} and~\ref{CH3+H2a}, have been previously studied, see the reported barriers in Table~\ref{tab:barrier}. Both reactions can only take place when a considerable barrier ($>44$~kJ/mol) is overcome, for which tunneling needs to be invoked to reach rate constants that are high enough for the reaction to be able to take place at the low temperatures in dense molecular clouds. The effect of tunneling can be accurately included by means of instanton theory \citep{Langer:1976, Miller:1975, Callan:1977, Rommel:2011, Richardson:2016}. Although \citet{Beyer:2016} indeed calculated instanton rate constants for the reaction \ce{CH3 + H2}, they provided only bimolecular rate constants, whereas unimolecular rate constants are those relevant for Langmuir-Hinshelwood type surface reactions \citep{Lamberts:2016,Meisner:2017}. Given the high activation barriers, about double as for the reaction \ce{H2 + OH -> H2O + H} \citep{Meisner:2017} and slightly above that of \ce{H + H2O2 -> H2 + HO2} \citep{Lamberts:2016}, relatively low rate constants are expected. While the quantitative calculations will be the topic of a future study, these two reactions are expected to show significant isotope effects and this explains also without further theory why the experiments with only \ce{D2} as a source of deuterium do not lead to a \ce{CD4} detection. 

\begin{table*}[t]
    \centering 
    \caption{Activation energies from predominantly theoretical chemical literature for the reactions~\ref{C+H}--\ref{CH3+H2a} in kJ/mol.}\label{tab:barrier}
    \begin{tabular}{lr@{ }llc}
    \hline
    Reaction &      & & $\Delta E_\text{act}$  &    Ref. \\
    \hline 
    (\ref{C+H})     & \ce{^3C + H} & \ce{-> CH}  & 0                             & [1] \\
    (\ref{CH+H})    & \ce{CH + H} & \ce{-> CH2}  & 0                             & [2] \\
    (\ref{CH2+H})   & \ce{^3CH2 + H} & \ce{-> CH3}  & 0                             & [3] \\
    (\ref{CH3+H})   & \ce{CH3 + H} & \ce{-> CH4}  & 0                             & [4] \\
    (\ref{C+H2})    & \ce{^3C + H2} &\ce{-> ^3CH2} & $27-30$ \tablenotemark{a}             & [2] \\
    (\ref{CH+H2})   & \ce{CH + H2} &\ce{-> CH3} & 0                             & [3] \\
    (\ref{C+H2a}) & \ce{CH + H} &\ce{-> ^3C + H2}  &  --\tablenotemark{b}           & [2] \\
    (\ref{CH+H2a}) & \ce{CH2 + H} &\ce{-> CH + H2}  & --\tablenotemark{c}             & [3] \\
    (\ref{CH2+H2a}) & \ce{^3CH2 + H2} &\ce{-> CH3 + H} & 49                         & [5] \\    
    (\ref{CH3+H2a}) & \ce{CH3 + H2} &\ce{-> CH4 + H} & 44                         & [6] \\    
    \hline
\end{tabular} 
\tablenotetext{a}{Cannot be trivially determined due to conical intersection and strong C--\ce{H2O} interaction, see Section~\ref{sec:CH2PES}} 
\tablenotetext{b}{The reaction is likely quenched in the \ce{CH2} ground state, see Section~\ref{sec:CH2PES} and \citep{Gamallo:2012}, {effectively changing the reaction to}~\ref{CH+H}}
\tablenotetext{c}{Reaction is determined by the relaxation of the \ce{CH3^*} intermediate see Section~\ref{sec:CH3PES}, and {thus leads to} reaction~\ref{CH2+H} }
\tablerefs{[1]~\citet{Qasim:2020a}; [2] \citet{Harding:1993,Harrevelt:2002,Gamallo:2012}; this work; [3] \citet{McIlroy:1993, Medvedev:2006, Gonzalez:2011}; [4] \citet{Duchovic:1985}; [5] \citet{Baskin:1974,Bauschlicher:1978}; [6] \citet{Li:2015, Beyer:2016}}
\end{table*}

\section{Astrochemical Implications and Conclusions}\label{sec:astroconc}

Currently, the main formation pathway of methane is thought to be through the reactions~\ref{C+H}-\ref{CH3+H}. Reactions~\ref{C+H2}, ~\ref{CH2+H2a}, and~\ref{CH3+H2a} are included in some astrochemical models, however, in most studies the initial guesses by \citep{Hasegawa:1993} are used, see for instance the current surface reactions in the KIDA database \citep{KIDA}. Reaction~\ref{CH+H2}, \ce{CH + H2 -> CH3}, is usually not included at all. For the interpretation of astronomical data, specifically those obtained by JWST in the nearby future, it is important for models to take into account that: 
\begin{enumerate}
  \setlength{\itemsep}{0pt}
  \setlength{\parskip}{0pt}
    \item The reaction \ce{C + H2 -> CH2} is unlikely to proceed \emph{via} a fully barrierless mechanism on water ices and an isotope effect is yet to be determined. 
    \item The reaction \ce{CH + H2 -> CH3} on the other hand is expected to take place readily and barrierlessly without an isotope effect.
    \item The abstraction reactions \ce{CH2 + H2 -> CH3 + H} and \ce{CH3 + H2 -> CH4 + H} are expected to take place \emph{via} a tunneling mechanism and a pronounced isotope effect is expected.
\end{enumerate}
The points above are summarized in Table~\ref{tab:barrier} and Fig.~\ref{fig:network}.

{The main finding is that \ce{H2} plays a role in the solid state formation of interstellar methane.} \ce{CH4} can be efficiently formed without invoking any H-atoms at all in our experiments, which is supported by our own calculations as well as theoretical results found in physical chemical literature. Thus, under physical conditions where \ce{H2} is much more abundant than H atoms and taking into account that \ce{H2} sticks to the surface at higher temperatures than H, methane formation from C atoms and \ce{H2} (and \ce{HD}, \ce{D2}) molecules is a reaction route that should be taken into account. {In the end, both \ce{H2} and H abundances as well as their respective reaction efficiencies with carbon atoms determine the relative impact of both mechanisms, for which dedicated modeling will be needed.} The finding that a \ce{C + H2} route also leads to methane formation has the following implications: 
\begin{enumerate}
  \setlength{\itemsep}{0pt}
  \setlength{\parskip}{0pt}    
    \item The formation of \ce{CH4} can take place at ‘higher’ temperatures, \emph{e.g.}, 20~K instead of 10~K, because of the stronger binding of \ce{H2} molecules to the ice surface,
    \item \ce{CH4} can be formed in the ice bulk through the interaction of entrapped \ce{H2} with \ce{CH_{n}} radicals obtained by dissociation of hydrocarbons from UV-photons or cosmic ray particles,
    \item Deuterium fractionation of methane is not only dictated by D/H ratios but also by {(a) the respective abundances of \ce{D2} and HD with respect to \ce{H2} on the surface and (b) an isotope effect is expected because of the presence or lack of a barrier, see for instance Fig.~\ref{fig:allnetwork},}
    \item {While \citet{Qasim:2020a} focused on confirming the atomic hydrogenation route of carbon to form \ce{CH4}, here we show that not only reactions with H, but also \ce{H2} chemistry overall should be fully incorporated into astrochemical models.} This is particularly true for models that include microscopic detail, despite the increase in computational cost.
    \end{enumerate}
    
\software{Matplotlib \citep{Matplotlib}, Numpy \citep{numpy}, Jupyter \citep{jupyter}, Chemshell \citep{Chemshell}, Molpro \citep{Molpro} }

\acknowledgments

T.L. is grateful for support from NWO via a VENI fellowship (722.017.008). G.F. acknowledges financial support from the Russian Ministry of Science and Higher Education via the State Assignment Contract FEUZ-2020-0038. This research benefited from the financial support from the Dutch Astrochemistry Network II (DANII). Further support includes a VICI grant of NWO (the Netherlands Organization for Scientific Research). Funding by NOVA (the Netherlands Research School for Astronomy) is acknowledged.



\appendix

\section{Reaction routes for the formation of methane isotopologues} \label{sec:triple}
\begin{figure*}
    \centering
    \includegraphics[width=10.5cm]{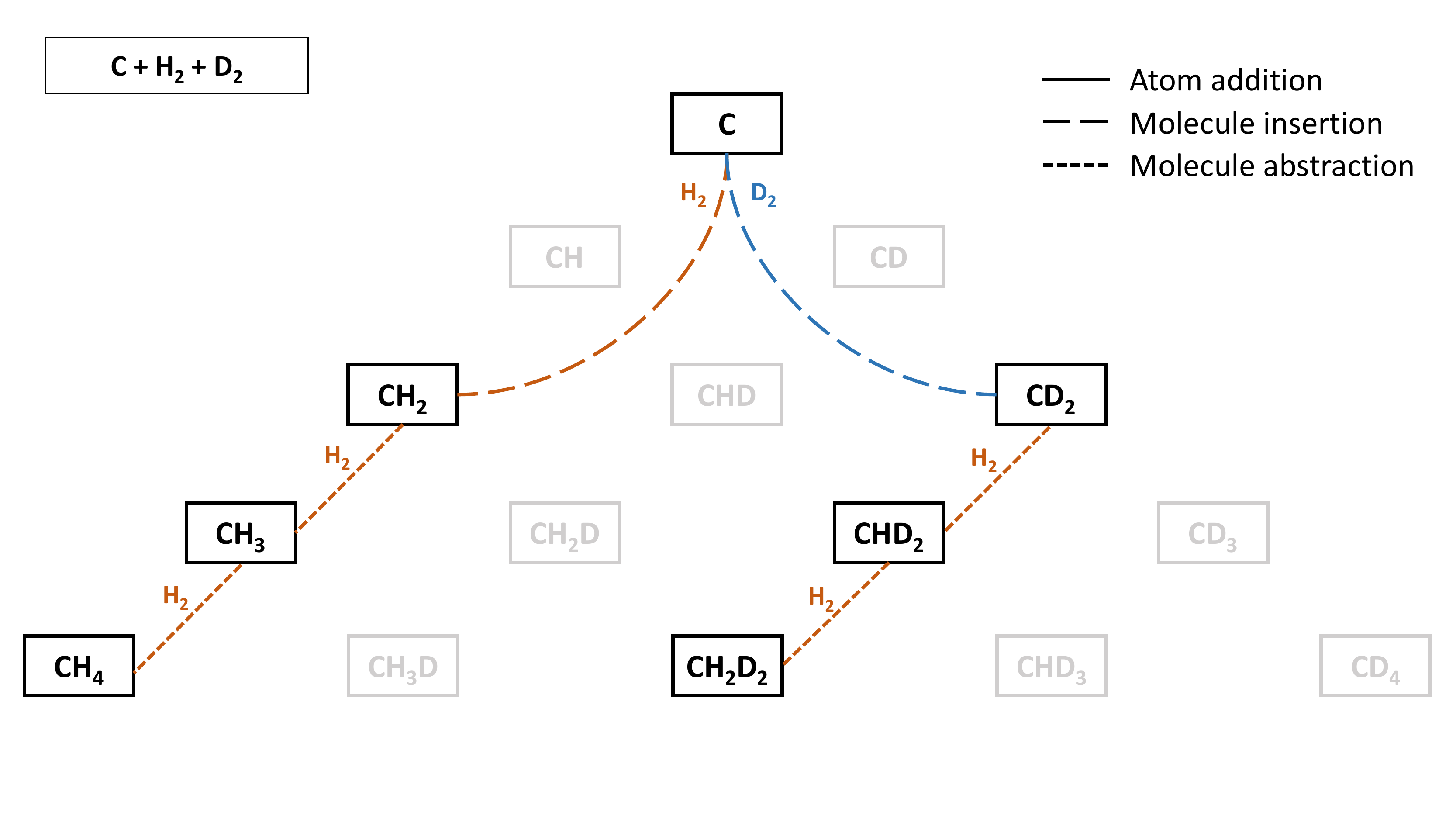}\\
    \includegraphics[width=10.5cm]{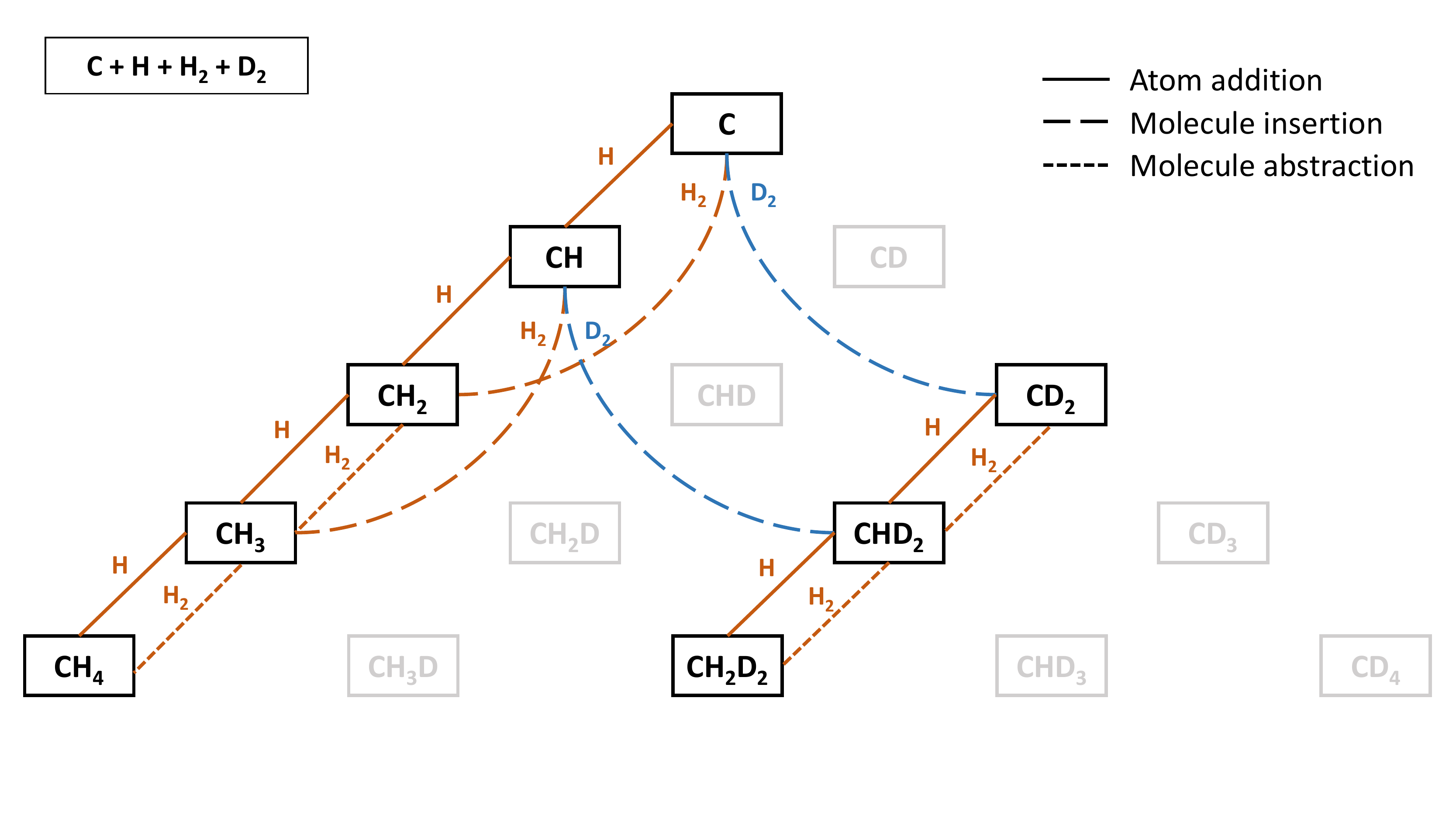}\\
    \includegraphics[width=10.5cm]{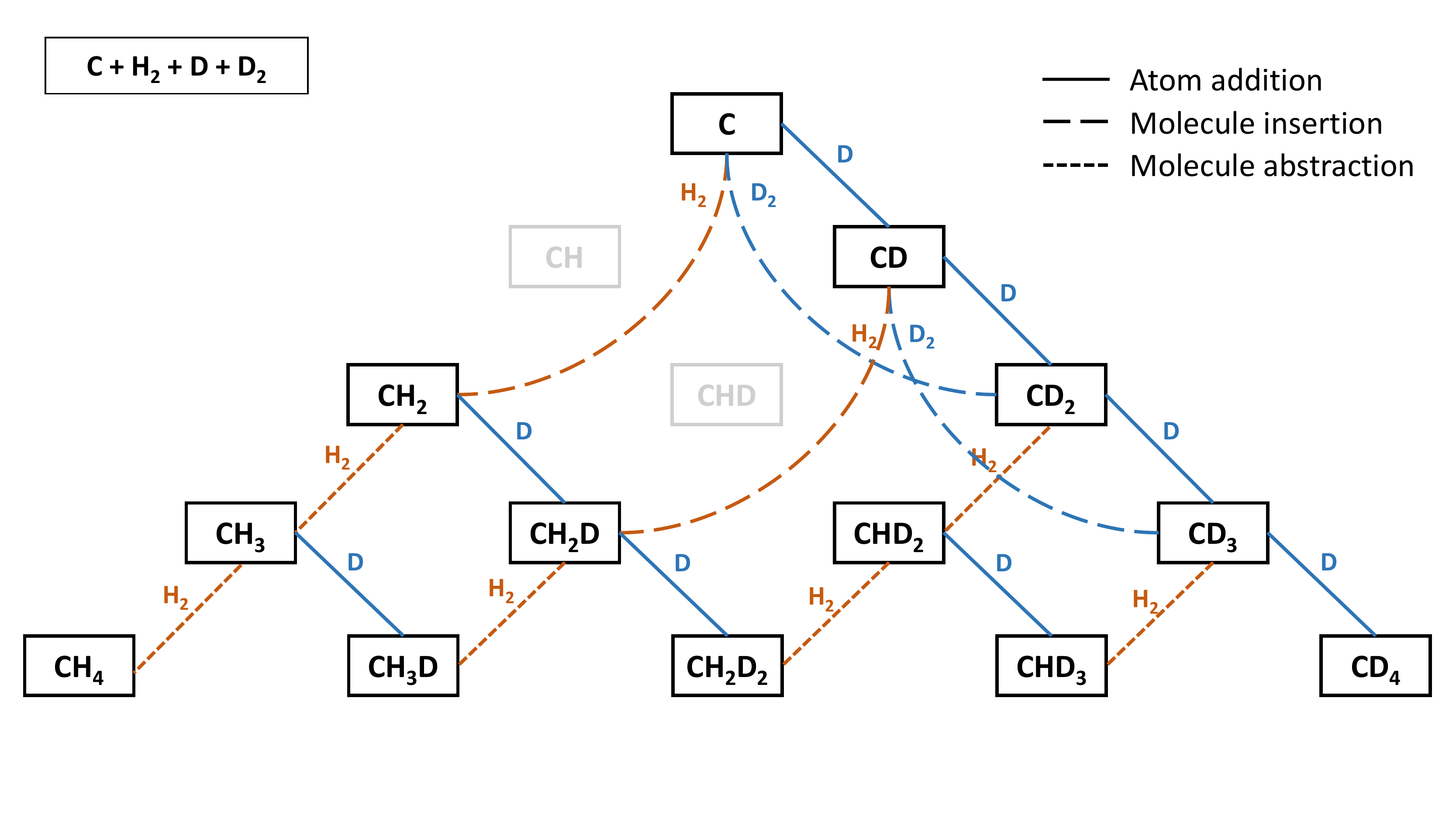}\\
    \caption{Reaction network for the hydrogenation and deuteration reactions leading to the formation of methane isotopologues. Analogously to Fig.~\ref{fig:network}, three types of reactions are considered: H or D addition, \ce{H2} or \ce{D2} insertion, and \ce{H2} abstraction.}\label{fig:allnetwork}
\end{figure*}

Below we explain the reaction pathways that lead to the formation of methane isotopologues of the form \ce{CH_{n}D_{4-n}} with $n=0-4$ in the experimental series $1-4$. Please note that the reactions considered barrierless are reactions~\ref{C+H}--~\ref{CH3+H} and~\ref{CH+H2} and these are expected to take place without an isotope effect. Reaction~\ref{C+H2} is likely possible both with \ce{H2} and \ce{D2}, although it is currently unclear whether the rate constant is determined by the change needed in electron configuration and/or a barrier on a water-rich surface. It is possible that the reaction is slower with \ce{D2}. Reactions~\ref{CH2+H2a} and~\ref{CH3+H2a} can only take place via tunneling and, given the high barrier, these reactions are expected to be very slow with \ce{D2}. {In Fig.~\ref{fig:allnetwork} three networks are depicted, analogously to Fig.~\ref{fig:network}, one for each deuteration experiment. As in Fig.~\ref{fig:network}, three types of reactions are considered, H/D atom addition, \ce{H2/D2} insertion, and \ce{H2} abstraction reactions. Note that we deliberately choose not to include \ce{D2} abstraction reactions, because of the high barrier.
Finally, we assume that the H or D atoms formed \emph{in situ} do not take part in subsequent reactions, but desorb instead. We base this on the argument of conservation of energy and momentum \citep{Koning:2013}.}

The experiments are discussed below in order of increasing complexity.

\subsection{Experiments 3 and 4: \ce{C + H2 and/or D2}}
The experiment (4A) with \ce{C + H2} leads clearly to the formation of \ce{CH4}, while an experiment (3A) with \ce{C + D2} does not lead to a \ce{CD4} detection. At the same time experiment 3B with \ce{C + H2 + D2} shows the formation of \ce{CH2D2} and \ce{CH4}, but not \ce{CD4}. This can be understood by considering the reactions presented in the top panel of Fig.~\ref{fig:allnetwork}. 

\subsection{Experiment 1: \ce{C + H + H2 + D2}}
Throughout the series for experiment 1 the only methane isotopologue species observed are \ce{CH4} and \ce{CH2D2}. The reaction network in Fig.~\ref{fig:allnetwork} rationalizes these two detections by showing that the energetically likely routes lead to these two species, while the formation of \ce{CHD3} or \ce{CH3D} would proceed through a \ce{+ D2} tunneling reaction, which {is unlikely to take place at laboratory timescales. Whether it would take place on interstellar timescales depends on the competition with diffusion and needs an astrochemical model for verification}. 

\subsection{Experiment 2: \ce{C + H2 + D + D2}}
In experimental series 2 we confirm the detection of \ce{CD4}, \ce{CH2D2}, and a tentative detection of \ce{CH4}. However, as can be seen from the reaction network in Fig.~\ref{fig:allnetwork}, both \ce{CHD3} and \ce{CH3D} could in principle be formed. We attribute the lack of detecting these species to: 
\begin{enumerate}
  \setlength{\itemsep}{0pt}
  \setlength{\parskip}{0pt}
    \item the existence of a fully barrierless formation pathway for both \ce{CD4} and \ce{CH2D2},
    \item the large bandstrength for \ce{CH4} allowing a detection of small amounts,
    \item for \ce{CHD3} and \ce{CH3D} there is the requirement of at least one reaction that involves tunneling of the type \ce{CH_nD_m + H2} which is in direct competition with a barrierless reaction of the type \ce{CH_nD_m + D}.
\end{enumerate}   

\section{Peak positions and detections}\label{sec:peaks}
In Table~\ref{tab:peaks} the observed peak positions are listed, along with their molecular assignment based on literature values. The last four columns indicate for which of the experiments a particular peak has been detected, with brackets indicating a weak feature. 

We confirm the results published by \citet{Qasim:2020a} who showed that (deuterated) methane is efficiently formed when carbon atoms react with H (D) on a water surface via experiments 1A and 2A. Furthermore, we detect \ce{H2CO} and \ce{CO2}. Formaldehyde is present as a product from the reaction between the carbon atom and water \citep{Hickson:2016} and is the topic of a future study \citep{Molpeceres:2021}. The presence of \ce{H2CO} further leads to the tentative detection of a \ce{CH3OH} feature in experiments 1A and 1D at 1015 cm$^{-1}$ as a result of the hydrogenation of formaldehyde \citep{Watanabe:2002, Fuchs:2009, Qasim:2018}. \ce{CO2} is a contaminant that arises from atomic carbon sources of this design \citep{Krasnokutski:2014, Qasim:2020b}. Note also the gas-phase \ce{CO2} bands around 2340 cm$^{-1}$. 

\begin{table*}[t]
    \centering 
    \caption{Summary of all detected peak positions, with the exception of water (\ce{H2O}: 3380 and 1660 cm$^{-1}$ and \ce{D2O}: 2440 and 1220 cm$^{-1}$ ). Experiments in brackets indicate a weak feature or tentative detection}\label{tab:peaks}
    \begin{tabular}{llcllll}
    \hline
    Peak pos. &   Molecule      & Ref.      & Detected in  & Detected in   & Detected in & Detected in \\
     (cm$^{-1}$)  &             &           & Exp. 1       & Exp. 2        & Exp. 3     & Exp. 4\\
    \hline  
    3007        & \ce{CH4}      & [2]       & 1A, 1B   & --             & --    & (4A) \\
    3000        & \ce{CH2D2}    & [3]       & (1B), 1C & --           & --    & -- \\
    2343        & \ce{CO2}      & {[4]}   & All       & All           & 3A, 3B & \tablenotemark{a} \\
    2277        & \ce{CH2D2}    & [3]       & 1C      & (2C)          & --    &  -- \\
    2250        & \ce{CD4}      & [5]       & --            & 2A, 2B        & --    &  -- \\
    2226        & \ce{CH2D2}    & [3]       & 1C            & --            & --    &  -- \\
    2152        & \ce{CO}       & [6]       & All           & All           & 3A, 3B & All \\
    2137        & \ce{CO}       & [6]       & All           & All           & 3A, 3B & All \\
    1717        & \ce{H2CO}     & [7]       & All          & All          & 3A, 3B & -- \\
    1666         & \ce{D2CO}      & {[8]}       & --   &  --  & -- & 4A, 4B \\
    1500        & \ce{H2CO}     & [7]       & All          & All          & 3A, 3B & -- \\
    1430        & \ce{CH2D2}    & [3]       & 1B, 1C    &  --           & --      & -- \\
    1303        & \ce{CH4}      & [1,2]       & 1A, 1B, (1C) & (2C)         & (3B) & 4A \\
    1250        & \ce{H2CO}     & [7]       & 1A             & --            & 3A, 3B &  -- \\
    1231        & \ce{CH2D2}    & [3]       & 1B, (1B)  & --            & --    & -- \\
    1102        & \ce{D2CO}      & {[8]}       & --   &  --  & -- & (4A), 4B \\
    1083        & \ce{CH2D2}    & [3]       & 1B,         & (2C)          & 3B & -- \\
    1028        & \ce{CH2D2}    & [3]       & 1B, (1B) & (2A), 2B, 2C & 3B & -- \\
    1015        & \ce{CH3OH}    & [9]       & 1A, (1B)      & --            & --   & -- \\
    993         & \ce{CD4}      & [5]       & (1C)          &  2A, 2B, (2C)  & -- & -- \\
    991         & \ce{D2CO}      & {[8]}       & --   &  --  & -- & 4A, 4B \\
    \hline
    \end{tabular}\\
\tablerefs{[1] \citet{Shimanouchi:1972} ; [2] \citet{Hagen:1983, Quattrocci:1992a}; [3] \citet{Quattrocci:1992b}; [4] \citet{Gerakines:1995} ; [5] \citet{Chapados:1972, Edling:1987}; [6] \citet{Schmitt:1989}; [7] \citet{Schutte:1996}; [8] \citet{Tso:1984, Nagaoka:2005}; [9] \citet{Qasim:2018}}    
\tablenotetext{a}{Overlaps with the \ce{D2O} stretch mode}
\end{table*}

\section{Potential energy surfaces for the \ce{C_{2v}} and \ce{C_s} symmetries} \label{sec:pes}

\begin{figure*}
    \includegraphics[width=8.2cm]{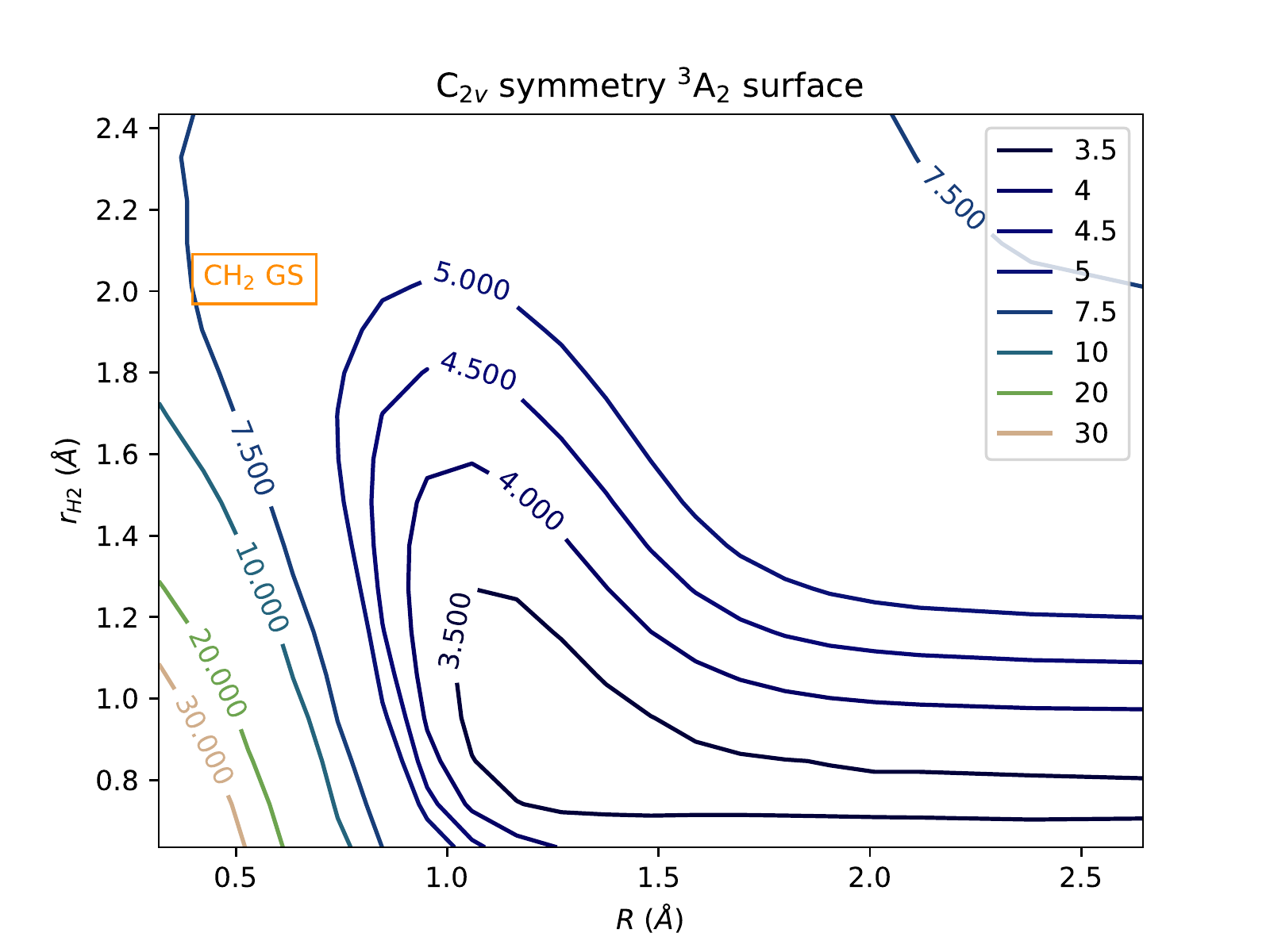}\hfill
    \includegraphics[width=8.2cm]{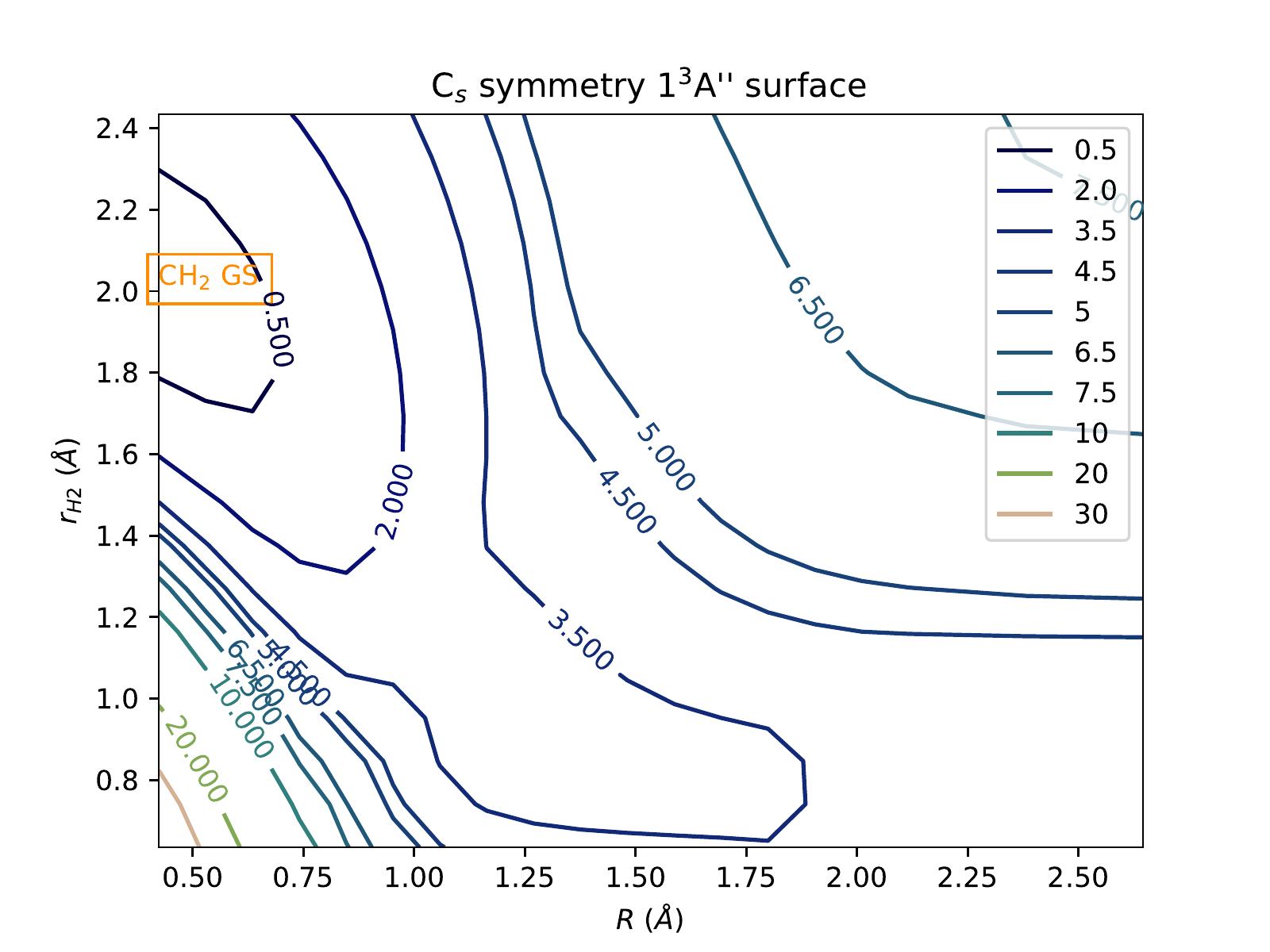}\\    \includegraphics[width=8.2cm]{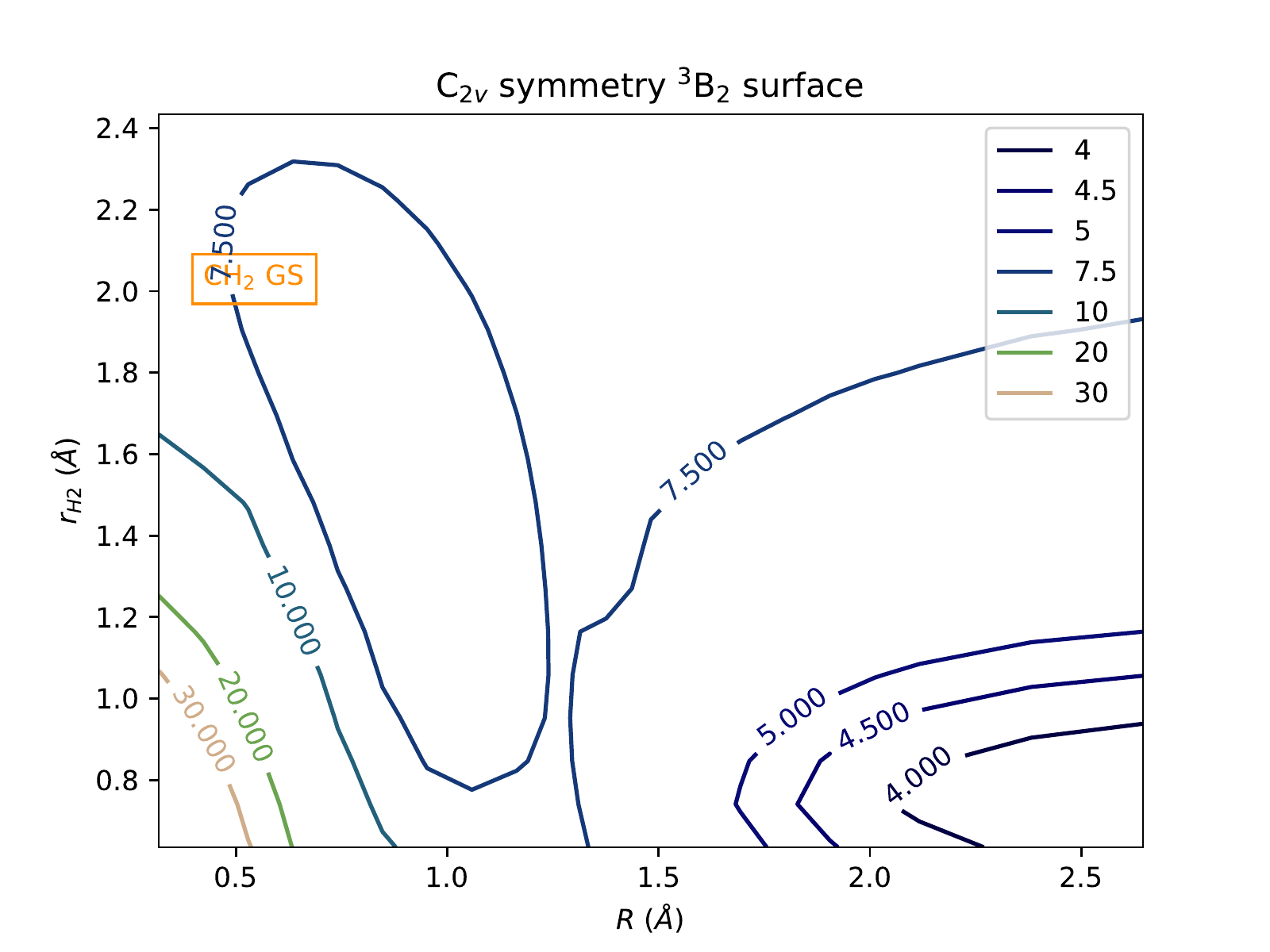}\hfill 
    \includegraphics[width=8.2cm]{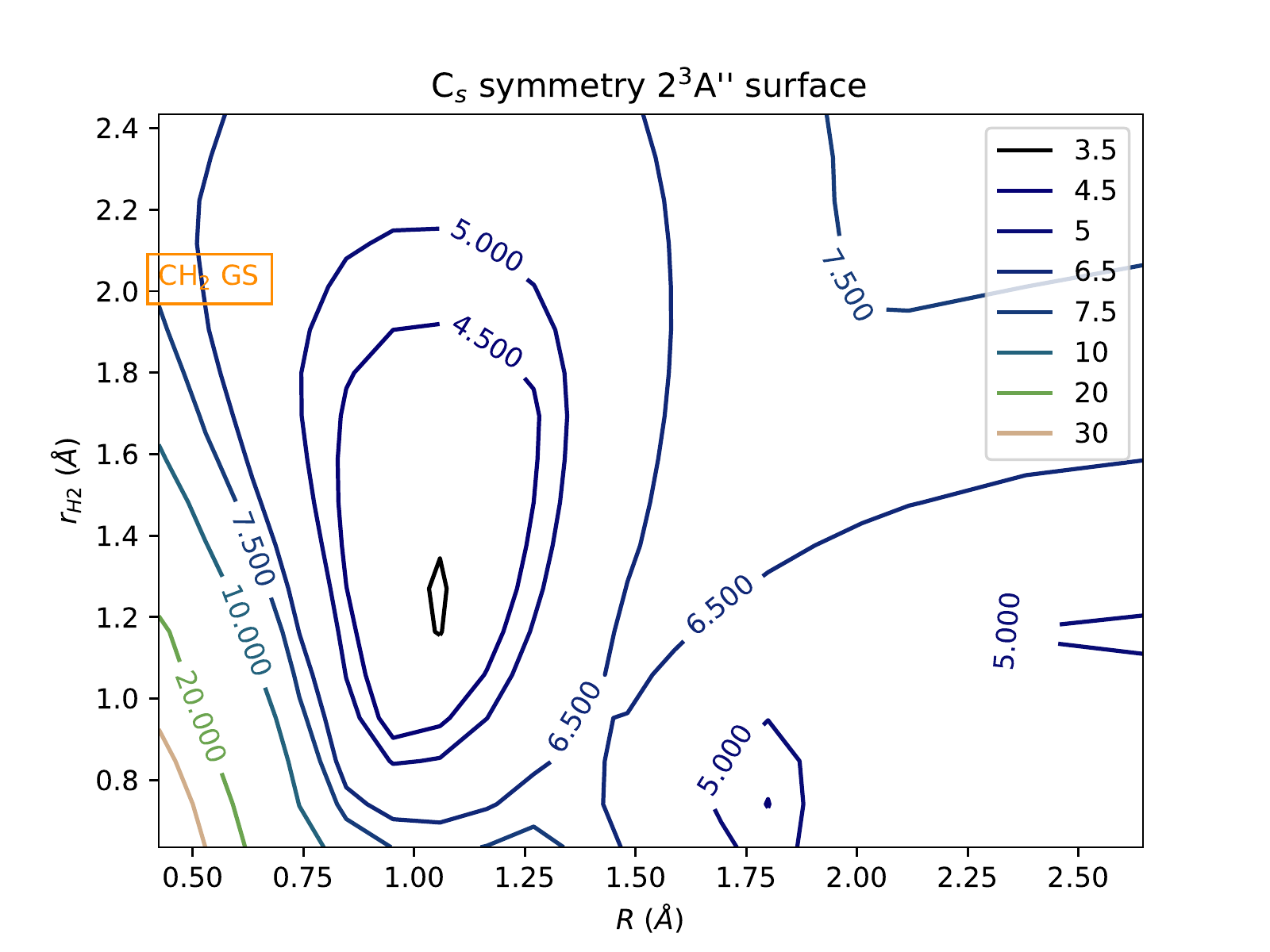}\\
    \includegraphics[width=8.2cm]{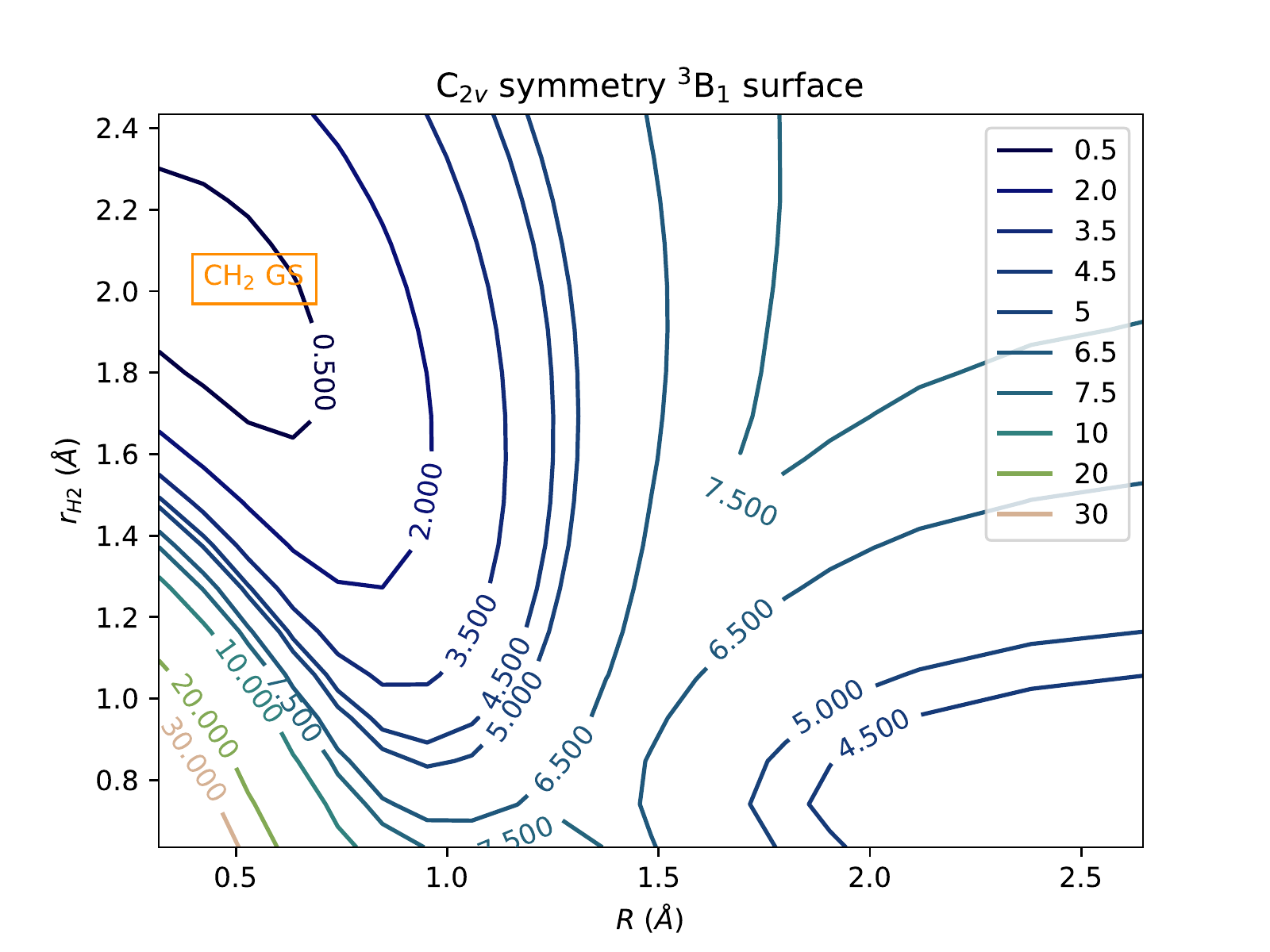}
    \caption{Potential energy cuts for the reaction \ce{C + H2} on three \ce{C_{2v}} (left column) and two \ce{C_s} symmetry (right column) surfaces.} \label{fig:pes}
\end{figure*}

Figure~\ref{fig:pes} shows the potential energy cuts for the reaction \ce{C + H2 -> CH2}, reaction~\ref{C+H2}, on both \ce{C_{2v}} and \ce{C_s} surfaces. Note that for both \ce{C_s} surfaces at $R>1.8$ $\AA$ we faced convergence issues. These values have been omitted. The position of the \ce{^3CH2} ground state is indicated in each figure.

\bibliography{bibCH2}{} 
\bibliographystyle{aasjournal}


\end{document}